# OPTIMIZING ONE FAIR DOCUMENT EXCHANGE PROTOCOL

## Abdullah M. Alaraj

Department of IT, Computer College, Qassim University, Saudi Arabia
arj@qu.edu.sa

## ABSTRACT

*This paper presents an efficient fair document exchange protocol. The exchange of the documents will be between two parties. The protocol is based on the verifiable and recoverable encryption of a document's key. This verifiable and recoverable encryption of the document's key will allow one party to verify the encrypted key. It will also ensure this party that the Semi Trusted Third Party will be able to recover the key if the other party misbehaves. The protocol also incorporates the concept of enforcing the honesty of one party. The proposed protocol consists of only three messages and is more efficient than related protocols.*

## 1. INTRODUCTION

Individuals and businesses are relying on the Internet for conducting different types of transactions. One of these transactions is the exchange of valuable documents (such as electronic payment and products) between the parties. That is, party A will exchange its valuable document for party B's valuable document. As an example of such an exchange, party A would like to buy an electronic product (e-product such as computer game) from party B. As parties using communication networks, they cannot send their documents at the same time. Rather, one party sends its document at a time. After receiving the document of the first party, the second party sends its document.

There are risks associated with such exchange. One of the most important risks is the case where party A sends its document to party B but the later either disappears before sending its document to party A or sends an incorrect document. Therefore, party A will be the loser in this scenario because the party who sends its document first will be at risk. This problem is known as the fairness problem. The fairness problem is solved using fair exchange protocols that ensure the fair exchange of documents between the parties involved. That is, fair exchange protocols will ensure that either both parties get each other's item or none do.

The contribution of this paper is that it applies the concept of enforcing the honesty of one party to the verifiable and recoverable encryption of a document's key proposed by Zhang et al [12]. The result of this application is a new optimized fair document exchange protocol as will be shown in the comparison in section 5.

The paper is organized as follows. Section 2 will be discussing the literature survey. Section 3 will present the new protocol. The analysis of the proposed protocol and comparison will be discussed in sections 4 and 5, respectively.

## 2. LITERATURE SURVEY

A number of fair exchange protocols have been proposed in the literature [1,2,3,4,7,9,10,11,12,14, 19]. These protocols are either based on a Trusted Third Party (TTP) or gradual exchange protocols. The gradual exchange protocols [10] allow the parties to exchange their items without involvement of any other party. The TTP-based protocols require a TTP to be involved. The involvement of the TTP can be either online such as in [7, 9, 17] or offline such as in [1, 2, 3, 4, 11, 12, 19]. The online TTP must be available during the exchange of items between parties because one of the parties (or all of the parties involved) will use it either for verification purposes or downloading items. The offline TTP will not be involved during the exchange of items between parties. Rather, it will be contacted in case one party misbehaves.

The fair exchange protocols can be used to exchange any two items between two (or more) parties. The items can be valuable documents, a document and payment, two digital signatures on a contract, and an email with a receipt. The focus in this paper is on fair exchange protocols that are for the exchange of two valuable documents between two parties.

Zhang et al [12] proposed a fair document exchange protocol between two parties A and B. The protocol is based on the verifiable and recoverable encryption of keys. Parties A and B will first exchange their encrypted documents in the first two messages. Then, the parties will exchange the decryption keys to decrypt the encrypted documents. If one party misbehaves, the offline STTP (Semi Trusted Third Party that will not collude with any party but may misbehave by itself) can be contacted to recover the key. To start the protocol, party A will send its encrypted document to party B. Party B will then verify the correctness of the encrypted document. If it is correct, then party B will send the following to party A: (a) its encrypted document, (b) verifiable and recoverable encryption of the key that encrypts the document, and (c) the authorization token. Party A will then verify the correctness of the encrypted document, authorization token and the encrypted key. If these verifications are correct, then it is safe for party A to send its decryption key to party B. Finally, once party B decrypted the document, it sends its decryption key to party A. If party B misbehaves by either sending an incorrect decryption key or not sending the decryption key to party A, then party A can contact the STTP to recover the decryption key.

Ray et al [7] proposed a fair exchange protocol for the exchange of documents (e.g. digital products and payments between customers and merchants). The protocol is based on cross validation theorem that states [7] "if a message is encrypted with the product key of two compatible keys and another message is encrypted with either of the two compatible keys and the two encrypted messages compare, then the two original unencrypted messages must also compare".

In the protocol, a merchant M exchanges a digital product for a payment from a customer C. Before the protocol starts, the merchant (M) needs to register with a trusted third party (TTP). The TTP generates the key pair $KM_1$ and $KM_1^{-1}$. The TTP then provides M with $KM_1$ and keeps $KM_1^{-1}$ with itself. C needs to have an account in a bank. The bank generates the key pairs $KC_1$ and $KC_1^{-1}$. The bank then provides C with $KC_1$ and keeps $KC_1^{-1}$ with itself. M needs to send the digital product, its description and its price to the TTP. The TTP encrypts the digital product using the key $KM_1$ and then advertises it on its website. C needs to download the encrypted digital product from the TTP.

The exchange part of Ray et al protocol [7] consists of four messages. C sends to M the first message that includes the purchase order and the payment that is encrypted with the product key of ($KC_1$ x $KC_2$). Then, M sends the second message to C. The second message includes the digital product that is encrypted with the product key of ($KM_1$ x $KM_2$). On receiving the second message, C compares the hash value of the encrypted digital product that was downloaded from the TTP with the hash value of the encrypted digital product that is included in the second message. If the two hash values are matched then C can be sure that the unencrypted digital products will be matched as well. After verifying that the two hashes are compared, C sends the third message to M. The third message includes the decryption key for the encrypted payment. Finally, M sends the fourth message to C which includes the decryption key of the encrypted digital product. If M misbehaves, C contacts the TTP for the recovery of the decryption key of the digital product.

Alaraj and Munro [1] proposed a fair exchange protocol for the exchange of two documents (the two documents can be a digital product and payment) between a customer and a merchant. Alaraj and Munro proposed a new design approach for the exchange. They call it enforcing the customer to be honest. The protocol works as follows. The merchant starts the protocol by sending the first message to the customer. The first message includes the merchant's document encrypted with a key. This key is also encrypted using a shared public key between the merchant and the TTP. On receiving the first message, the customer will verify the encrypted document and the encrypted key. If they are correctly verified then the customer will send the second message to the merchant. The second message includes the customer's document encrypted with a key that was sent to the customer by the merchant in the first message. On receiving the second message, the merchant will use the key that it already has to decrypt the customer's document. When the document is decrypted correctly, the merchant will send the decryption key to the customer. If the merchant refuses to send the decryption key, the customer can contact the TTP to send the decryption key to the customer. This approach is called enforcing the customer to be honest because the customer can not cheat by sending an incorrect document because they are going to encrypt their document using a key that the merchant already has. Using this approach, Alaraj and Munro [1] were able to propose a fair exchange protocol using only three messages.

Alaraj and Munro [3] proposed a protocol that is similar to the protocol in [1]. The difference is that the merchant is the one who is enforced to be honest in [3].

The design approach of most of the protocols proposed in the literature, apart from Alaraj and Munro [1, 3], is to include at least four messages in the exchange protocol. The first two messages are for the exchange of the encrypted items between the participating parties. The last two messages are for the exchange of decryption keys to decrypt the items received in the first two messages. The design approach of Alaraj and Munro [1, 3] is to have only three messages in the protocol. The first message includes the encrypted item of the first party. The other party will be able to verify it and if it is correctly verified then they will send the second message to the first party. The second message includes the encrypted item of the second party but the first party will be able to decrypt it as it is encrypted with a key that the first party already has. Therefore, the second party has to send a correct item in order to receive the decryption key of the first party's item in the third message. Therefore, the design approach of Alaraj and Munro protocols [1, 3] is based on the exchange of an item (i.e. that is included in the second message) for a decryption key (i.e. that is included in the third message). The result is to have more efficient protocol that includes only three messages.

The proposed protocol in this paper uses the concept of having one party to be enforced to be honest to reduce the number of messages. Moreover, the concept of verifiable and recoverable encryption of keys is also used. Therefore, more efficient protocol is proposed.

## 3. THE DOCUMENT EXCHANGE PROTOCOL

### 3.1 Notations

The following represents the notations used in the proposed protocol:

- $P_a$: party a
- $P_b$: party b
- STTP: Semi Trusted Third Party is a party neither $P_a$ nor $P_b$. STTP will not collude with any other party but may misbehave by itself
- h(X): a strong-collision-resistant one-way hash function, such as SHA-1 [13]
- $pk_x = (e_x, n_x)$: RSA Public Key [16] of the party $x$, where $n_x$ is a public RSA modulus and $e_x$ is a public exponent
- $sk_x = (d_x, n_x)$: RSA Private Key [16] of the party x, where $n_x$ is a public RSA modulus and $d_x$ is a private exponent
- $D_x$: the document of party x
- $k_x$: a symmetric key that will be used for encryption and decryption of a document
- $C_{-bt}$: the certificate for the shared public key between $P_b$ and the STTP. $C_{-bt}$ is issued by the STTP. A standard X.509 certificate [15] can be used to implement $C_{-bt}$
- enc.$pk_x$(Y): an RSA [16] encryption of Y using the public key $pk_x$ ($e_x$, $n_x$). The encryption of Y is computed as follows. enc.$pk_x$(Y) = $Y^{ex}$ mod $n_x$
- enc.$sk_x$(Z): an RSA [16] decryption of Z using the private key $sk_x$ ($d_x$, $n_x$). The decryption of Z is computed as follows. enc.$sk_x$(Z) = $Z^{dx}$ mod $n_x$
- enc.$k_x$(Y) : encryption of Y using a symmetric key $k_x$ ($k_x$ can be used for decrypting enc.$k_x$(Y))
- Sig.$_a$ (X): the RSA digital signature [16] of the party $a$ on X. The digital signature of party $a$ on X is computed by encrypting the hash value of X using the private key $sk_a$ ($d_a$, $n_a$). This is computed as follows. Sig.$_a$ (X) = $(h(x))^{da}$ mod $n_a$
- A → B: X: A sends message X to B
- X + Y: concatenation of X and Y
- $heD_x$: hash value of encrypted $D_x$ using $k_x$

## 3.2 Assumptions

The following represents the assumptions made for the proposed protocol:

- Each party ($P_b$, $P_a$ and STTP) has its own public and private keys.
  - The STTP's public key is denoted as $pk_t = (e_t, n_t)$ and its corresponding private key is denoted as $sk_t = (d_t, n_t)$.
  - $P_b$'s public key is denoted as $pk_b = (e_b, n_b)$ and its corresponding private key is denoted as $sk_b = (d_b, n_b)$.
  - $P_a$'s public key is denoted as $pk_a = (e_a, n_a)$ and its corresponding private key is denoted as $sk_a = (d_a, n_a)$.
- $P_b$ has a RSA-based public-key certificate $C_{-bt} = (P_b, pk_{bt}, W_{bt}, Sig._t)$ issued by STTP [12]. The content of $C_{-bt}$ is described as follows.
  - $P_b$ in $C_{-bt}$ is $P_b$'s identity to make $C_{-bt}$ valid only for $P_b$.
  - The public key $pk_{bt}$ and its associated private key $sk_{bt}$ are denoted as $pk_{bt} = (e_{bt}, n_{bt})$ and $sk_{bt} = (d_{bt}, n_{bt})$, respectively, where $n_{bt}$ is a product of two distinct large primes chosen randomly by STTP. This pair of keys needs to be produced in relation to $P_b$'s public key $pk_b = (e_b, n_b)$ so that $e_{bt} = e_b$ and $n_{bt} > n_b$ [12]. STTP does not allow any other party, including $P_b$, to know $sk_{bt}$, and it sends only $C_{-bt}$ to $P_b$. One $C_{-bt}$ certificate will be issued for $P_b$, and $P_b$ can use $C_{-bt}$ for as many document exchanges as $P_b$ wishes [12]
  - $W_{bt}$ in $C_{-bt}$ is defined as $W_{bt} = (h(sk_t + pk_{bt})^{-1} * d_{bt})$ mod $n_{bt}$, where $sk_t$ is STTP's private key, and $h(sk_t + pk_{bt})^{-1}$ is the multiplicative inverse of $h(sk_t + pk_{bt})$ modulo $n_{bt}$,
    i.e. $h(sk_t + pk_{bt})^{-1} h(sk_t + pk_{bt})$ mod $n_{bt} = 1$.

$W_{bt}$ is included in $C_{bt}$ in order to eliminate the need for STTP to store and safe-keep private key $sk_{bt}$ [12]. Therefore, STTP will compute it from $W_{bt}$, i.e. $d_{bt} = (h(sk_t + pk_{bt}) W_{bt}) \bmod n_{bt}$

o   $Sig_{t}$ in $C_{bt}$ is STTP's RSA signature on $h(P_b, pk_{bt}, W_{bt})$, i.e. $Sig_{t}$=enc.sk$_t$(h(P$_b$ + pk$_{bt}$ + W$_{bt}$))

- The following is known to $P_b$ before the exchange protocol is executed:
  o   $heD_a = h(enc.k_a(D_a))$ which is the hash value of encrypted $D_a$ with $k_a$
- The following is known to $P_a$ before the exchange protocol is executed:
  o   $ek_b = enc.pk_b(k_b)$ which is the encryption of $k_b$ with the public key of $P_b$

## 3.3 Protocol description

Semi Trusted Third Party (STTP) will be used in the proposed protocol. The STTP may misbehave but it will not collude with any other party involved in the exchange [18].

The idea of the proposed protocol is to have one party ($P_b$) sends its first message to the other party ($P_a$). The first message includes the encrypted document, verifiable and recoverable encryption of $P_b$'s key (this key is used to encrypt $P_b$'s document) and the authorization token. The verifiable and recoverable encryption of $P_b$'s key allows $P_a$ to verify it and if it is correct then $P_a$ can be sure that STTP will be able to recover the key in case $P_b$ does not sends it i.e. if $P_b$ misbehaves. So, when $P_a$ verifies this verifiable and recoverable encryption correctly then $P_a$ will send its message that contains its encrypted document using a key that was sent to $P_a$ by $P_b$. Then, $P_a$ will wait for the third message from $P_b$ that includes the decryption key for the encrypted document received in the first message. If $P_b$ did not send the third message then $P_a$ will contact STTP to recover the key. The STTP will verify the authorization token generated by $P_b$ to make sure that $P_a$ provided what $P_b$ wants.

Therefore, for $P_b$ to produce this verifiable and recoverable encryption of $P_b$'s key $k_b$, $P_b$ chooses a large prime $r_b$ relatively prime to $n_b$ in $P_b$'s public key $pk_b=(e_b, n_b)$ and then computes the following [12]:

$X_b= r_b*k_b$, where chosen $r_b$ needs to ensure that $x_b <n_b$
$Y_b= r_b^{eb} \bmod (n_b * n_{bt})$, with key $pk_{bt} =(e_{bt}, n_{bt})$ and $n_b<n_{bt}$
$Z_b= k_b^{eb} \bmod (n_b * n_b)$

$X_b$, $Y_b$ and $Z_b$ form the verifiable and recoverable encryption of $P_b$'s key $k_b$. Note that $Y_b$ can be decrypted using either $sk_b$ or $sk_{bt}$ [7]. Therefore, either $P_b$ or STTP can recover $r_b$.
The $P_b$'s authorization token will be defined by $P_b$. $P_b$'s authorization token represents $P_b$'s RSA signature on $h(C.bt+Y_b+Y_a+P_a)$ [12]. That is, Sb= sk$_b$(h(C.bt + Y$_b$ + Y$_a$ + P$_a$)),where:

$Y_a = h(enc.ka(D_a))$, this $Y_a$ is specified by $P_b$.

The authorization $S_b$ represents $P_b$'s conditional authorization stating that STTP can recover $r_b$ from $Y_b$ (which will enable $P_a$ to derive $k_b$ from $X_b$) if and only if $P_a$ provides an item "i.e. enc.ka($D_a$)" for STTP such that $h(enc.ka(D_a))=Y_a$. STTP will verify this $S_b$ and if it is correct then STTP can be sure that this "enc.ka($D_a$)" is the one that $P_b$ is looking for.
Therefore, the verifiable and recoverable encryption of key "$k_b$" will be generated by $P_b$, it will be verified by $P_a$, and it will be recovered by STTP.

## 3.4 Exchange Protocol

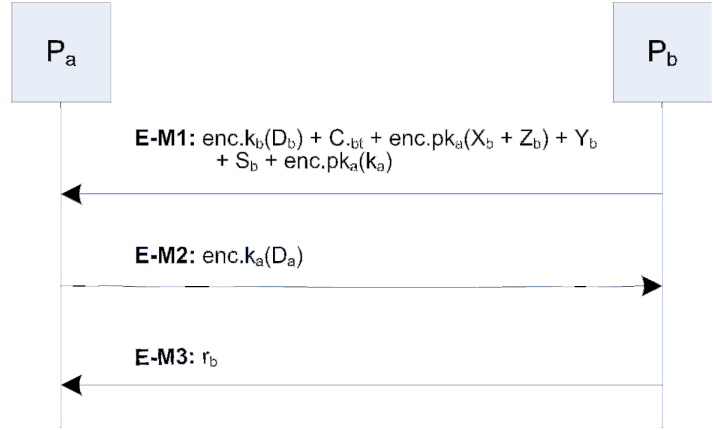

Figure 1: Exchange phase of the protocol

$P_b$ will start the exchange protocol by sending the first message E-M1 to $P_a$. The contents of E-M1 are as follows:

**E-M1: $P_b \rightarrow P_a$:** $enc.k_b(D_b) + C._{bt} + enc.pk_a(X_b + Z_b) + Y_b + S_b + enc.pk_a(k_a)$

The description of the contents of E-M1 is as follows:

- $enc.k_b(D_b)$ is the encryption of $P_b$'s document $D_b$ using $k_b$
- $C._{bt}$ is RSA-based public-key certificate that is discussed in section 3.2
- $enc.pk_a(X_b + Z_b)$ is the encryption of $X_b$ and $Z_b$ using $P_a$'s public key $pk_a$.
- $Y_b$
- $S_b$
- $enc.pk_a(k_a)$ is the encryption of ka using the public key of $P_a$. $k_a$ will later be used by $P_a$ to encrypt its document $D_a$. $k_a$ is chosen by $P_b$ and will be sent to $P_a$ to use it for encrypting its document $D_a$

On receiving the first message (E-M1), $P_a$ will make the following verifications [12]:

1. Verifying the correctness of $S_b$. This is done by decrypting $S_b$ using $P_b$' public key $pk_b$ to get the hash value included in the signature. Then, computing the hash value of $(C._{bt}+Y_b+Y_a+P_a)$. If the two hash values match then $S_b$ is correct.
2. Verifying the correctness of $C._{bt} = (P_b, pk_{bt}, W_{bt}, Sig._t)$ by decrypting $Sig._t$ using STTP's public key $pk_t$ to get the hash value included in the signature. Then, computing the hash value of $(P_b, pk_{bt}, W_{bt})$. If the two hash values match then $C._{bt}$ is correct.
3. Compute the hash value of $enc.k_b(D_b)$ and then compare it with $heD_b$. If the two hash values match then $P_a$ is sure that the encrypted $D_b$ is the one that $P_a$ is looking for
4. Confirm that $X_b < n_b$, and $Z_b \bmod n_b = enc.pk_b(k_b)$. It is assumed that $enc.pk_b(k_b)$ is known to $P_a$ (section 3.2)
5. Confirm that $X_b^{eb} \bmod n_b = (Y_b * enc.pk_b(k_b)) \bmod n_b$
6. Confirm that $X_b^{eb} \bmod n_{bt} = (Y_b * Z_b) \bmod n_{bt}$

If all verifications above are correct then it is secure for $P_a$ to send its document $D_a$ that is encrypted with a key that $P_b$ already has. Otherwise, $P_a$ terminated the protocol. So, if all verifications are correct then $P_a$ will send the second message (E-M2) to $P_b$ as follows:

**E-M2: $P_a \rightarrow P_b$:** enc.$k_a(D_a)$

The description of the contents of E-M2 is as follows:

- enc.$k_a(D_a)$ is the encryption of $P_a$'s document using $k_a$. $k_a$ was sent to $P_a$ in E-M1

On receiving E-M2, $P_b$ will do the following:

- Compute the hash value of enc.$k_a(D_a)$ then compare it with he$D_a$ (it is assumed that he$D_a$ is known to $P_b$ , section 3.2)

If the above verification is correct then Pb will decrypt $D_a$ using $k_a$ (note that, $k_a$ is already known to $P_b$). Then, $P_b$ will send E-M3 to $P_a$ as follows:

**E-M3: $P_b \rightarrow P_a$:** $r_b$

On receiving E-M3, $P_a$ will compute $k_b$ as follows:
$k_b = X_b/r_b$

Then, $P_a$ will use the key $k_b$ to decrypt enc.$k_b(D_b)$ to retrieve $D_b$.
At this step, both $P_a$ and $P_b$ have each other's documents i.e. they have fairly exchanged their documents.

### 3.5 Dispute Resolution Protocol (Key recovery protocol)

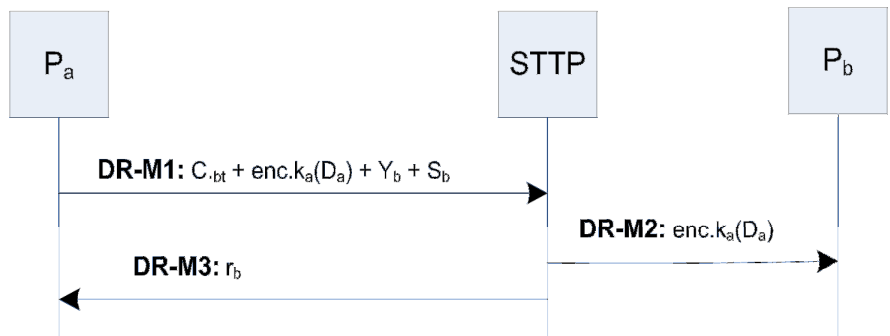

Figure 2: Dispute Resolution Phase of the Protocol

In the case of dispute (where $P_b$ misbehaves by either sending incorrect E-M3 or not sending E-M3 at all), $P_a$ will initiate the dispute resolution protocol by sending the message DR-M1 to the STTP as follows.

**DR-M1: $P_a \rightarrow$ STTP**: $C_{-bt}$ + enc.$k_a(D_a)$ + $Y_b$ + $S_b$

On receiving the message DR-M1 from $P_a$, STTP will do the following verifications:

1. Verifying the correctness of $S_b$. This is done by decrypting $S_b$ using $P_b$' public key $pk_b$ to get the hash value included in the signature. Then, computing the hash value of $(C_{.bt}+Y_b+Y_a+P_a)$. If the two hash values match then $S_b$ is correct.
2. Verifying the correctness of $C_{.bt} = (P_b, pk_{bt}, W_{bt}, Sig_{.t})$ by decrypting $Sig_{.t}$ using STTP's public key $pk_t$ to get the hash value included in the signature. Then, computing the hash value of $(P_b, pk_{bt}, W_{bt})$. If the two hash values match then $C_{.bt}$ is correct
3. Compute the hash value of $enc.k_a(D_a)$ and then compare it with $Y_a$ ($Y_a$ includes the hash value of $enc.k_a(D_a)$).

If any of the verifications above is incorrect then STTP will send an error message to $P_a$. Otherwise, if all verifications are correct then STTP will calculate $r_b$ from $Y_b$. Therefore, STTP needs to decrypt $Y_b$ using the shared private key i.e. $sk_{bt}$. So, STTP needs first to retrieve $sk_{bt}$ from $C_{.bt}$ as discussed in section 3.2. After decrypting $Y_b$ and getting $r_b$ from it, STTP will send the following two messages.

**DR-M2: STTP → $P_b$:** $enc.k_a(D_a)$

On receiving DR-M2 from STTP, $P_b$ will compute the hash value of $enc.k_a(D_a)$ then compare it with $heD_a$. If the two hash values match then $P_b$ will get $D_a$ by decrypting $enc.k_a(D_a)$ using $k_a$ that $P_b$ already has.

**DR-M3: STTP → $P_a$:** $r_b$

On receiving DR-M3 from STTP, $P_a$ will compute $k_b$ as follows:
$k_b = X_b/r_b$

Then, $P_a$ will use the key $k_b$ to decrypt $enc.k_b(D_b)$ to retrieve $D_b$.

At this step, both $P_a$ and $P_b$ have each other's items and hence the fairness is ensured.

# 4. ANALYSIS

The analysis of the security of the verifiable and recoverable encryption of $P_b$'s key $k_b$ is the same analysis conducted in [12]. Therefore, readers are referred to Zhang et al [12].

The following discusses all scenarios of the protocol's messages E-M1, EM2, E-M3 and DR-M1.

All possible scenarios of E-M1 will be studied as follows.

- $P_b$ sends incorrect E-M1 to $P_a$. If so, $P_a$ will find that E-M1 is incorrect when $P_a$ makes the verifications (these verifications discussed in sections 3.4). So, if E-M1 is incorrect then $P_a$ will not send E-M2 to $P_b$.
- $P_b$ sends correct E-M1 to $P_a$. After $P_a$ makes sure that E-M1 is correct by applying the verifications (these verifications discussed in sections 3.4) it is $P_a$'s choice to complete the exchange by sending E-M2 to $P_b$. However, if $P_a$ decides to complete the exchange then $P_a$ is enforced to be honest i.e. $P_a$ has to send correct E-M2 to be able to receive E-M3 from $P_b$.

All possible scenarios of E-M2 will be studied as follows.

- $P_a$ sends to $P_b$ in E-M2: $enc.k_a(D_a)$ where $k_a$ used is the key sent to $P_a$ by $P_b$ in E-M1. So, $P_b$ will first decrypt the message to get $D_a$ and then send $r_b$ to $P_a$ in E-M3
- $P_a$ sends to $P_b$ in E-M2: $enc.k(D_a)$ where k used is not the one sent to $P_a$ in E-M1. So, $P_b$ will not send E-M3 to $P_a$ i.e. $P_b$ will not send $r_b$
- $P_a$ does not send E-M2 to $P_b$ at all. So, $P_b$ will not send E-M3 to $P_a$ i.e. $P_b$ will not send $r_b$
- $P_a$ sends incorrect $D_a$ encrypted with $k_a$. So, $P_b$ will not send E-M3 to $P_a$ i.e. $P_b$ will not send $r_b$
- $P_a$ sends incorrect $D_a$ encrypted with k i.e. incorrect key. So, $P_b$ will not send E-M3 to $P_a$ i.e. $P_b$ will not send $r_b$

All scenarios of E-M3 will be studied as follows.

- $P_b$ sends correct $r_b$. So, $P_a$ will use it to decrypt $P_b$'s document and the exchange protocol will be completed fairly.
- $P_b$ sends incorrect $r_b$. So, $P_a$ will contact the STTP to recover $r_b$.
- $P_b$ did not send $r_b$ at all i.e. $P_b$ received correct E-M2 but did not send E-M3. So, $P_a$ will contact the STTP to recover $r_b$.

Therefore, from the previous scenarios it is clear that the fairness is ensued for both $P_a$ and $P_b$ either through the exchange phase of the protocol or through the dispute resolution phase.

All scenarios of DR-M1 will be studied as follows.

- $P_a$ sends correct DR-M1 to STTP. So, STTP will make the necessary verifications (i.e. verifications discussed in section 3.5) then STTP will send DR-M2 to $P_b$ and DR-M3 to $P_a$
- $P_a$ sends incorrect DR-M1 to STTP. So, STTP will make the necessary verifications (i.e. verifications discussed in section 3.5) then STTP will send an abort message to $P_a$.

Therefore, if $P_b$ misbehaves by not sending E-M3 or by sending incorrect E-M3 then the fairness can be ensured by allowing $P_a$ to send a correct DR-M1 to STTP. STTP will then ensure fairness for both $P_b$ and $P_a$ by sending DR-M2 and DR-M3, respectively.

If $P_a$ misbehaves by contacting STTP (i.e. by sending DR-M1) after receiving E-M1 i.e. before sending E-M2 to $P_b$, then STTP will verify $P_a$'s request. If STTP finds that DR-M1 is not correct then STTP will reject $P_a$'s request. If however STTP finds that DR-M1 is correct then STTP will send DR-M2 to $P_b$ and DR-M3 to $P_a$ to ensure fairness for both parties. Therefore, $P_a$ will not gain any advantage over $P_b$.

STTP is not able to get the documents $D_a$ and $D_b$ because an encrypted $D_a$ will be sent to it in DR-M1. STTP does not have the key to decrypt it. Rather, STTP will use it to verify if $P_a$ sent what $P_b$ is looking for. $D_b$ is not sent to STTP at all. Therefore, STTP will not be able to get $D_a$ and $D_b$. Hence, it is Semi Trusted Third Party.

Non-repudiation can be assured in the proposed protocol by having the signatures of parties $P_b$ and $P_a$ on their items to be included in messages E-M1 and E-M2.

# 5. COMPARISONS

In this section, the proposed protocol will be compared against the relevant protocols in the literature. That is, the proposed protocol will be compared against protocols in the literature, which are for the exchange of two documents (two documents or a document and payment) and involve an off-line or on-line TTP or STTP. The proposed protocol will be compared against Zhang et al protocol [12], Ray et al protocol [7], Alaraj and Munro protocol [1].

The protocols will be compared against the following criteria: number of messages in the exchange phase, number of messages in dispute phase, number of encryptions and decryptions in the exchange phase, number of symmetric encryptions in the exchange phase, and whether both parties involved in dispute resolution phase i.e. does the STTP need to contact both parties to verify the dispute request.

The number of messages in the exchange phase of ECH protocol and our protocol is 3 whereas it is 4 messages in both Zhang and Ray protocols. The number of messages in the dispute resolution phase is almost the same for all protocols. The number of RSA encryptions and decryptions for our protocol is 13 whereas it is 16 for Zhang et al protocol [12]. This shows how the idea of enforcing the honesty of one party introduced in ECH protocol helped in reducing the number of messages and the number of RSA encryptions and decryptions of Zhang et al protocol [12]. The application of enforcing the honesty of a party to Zhang et al protocol [12] is the main focus of this paper.

It is worth mentioning that Zhang et al's protocol [12] is better in that it does not require the document of party $P_a$ to be sent to the STTP in the dispute resolution phase whereas our protocol requires the party $P_a$ to send its encrypted document "enc.$k_a(D_a)$" to the STTP in the dispute resolution phase. However, this does not mean that the STTP will be able to decrypt the document because STTP does not have the key $k_a$. Rather, it uses it for the verification purposes.

Table 1 presents all the comparisons between our protocol and other relevant protocols in the literature.

Table 1: Comparison between our protocol and other protocols

|  | Zhang [12] | Ray [7] | ECH [1] | Our Protocol |
|---|---|---|---|---|
| Number of messages in exchange phase | 4 | 4 | 3 | 3 |
| Number of messages in dispute phase | 3 | 3 to 5 | 3 | 3 |
| Number of RSA encryptions and decryptions in exchange phase | 16 | 27 | 12 | 13 |
| Number of symmetric encryptions and decryptions in exchange phase | 4 | 0 | 4 | 4 |
| Both parties are involved in dispute resolution | No | Yes | No | No |

## 6. CONCLUSION

We have proposed an improved protocol for fairly exchanging two valuable documents between two parties. The proposed protocol uses offline Semi Trusted Third Party (STTP) that will only be contacted if one party misbehaved. The protocol is based on applying the idea of enforcing the honesty of one party to the method of verifiable and recoverable encryption of keys. The outcome of this application is a more efficient fair document exchange protocol. Only three messages are required to exchange the valuable documents between the two parties. Additionally, the number of modular exponentiations is less in our protocol compared to the protocols based on verifiable and recoverable encryption of keys.

A future work will include formally evaluating the protocol and implementing it.